\documentclass[jcp,aps,twocolumn,superscriptaddress,floatfix,showpacs]{revtex4}

\newcommand{\xp}{x^{\prime}}
\newcommand{\wfx}{\psi(x)}
\newcommand{\wfxp}{\psi(x^{\prime})}

\newcommand{\triple}{\mathrm{C}\equiv \mathrm{C}}
\newcommand{\single}{\mathrm{C}-\mathrm{C}}
\newcommand{\ch}{\mathrm{C}-\mathrm{H}}
\newcommand{\form}{\mathrm{C}_N \mathrm{H}_2}
\newcommand{\formmin}{\mathrm{C}_N \mathrm{H}_2^{-}}
\newcommand{\formpl}{\mathrm{C}_N \mathrm{H}_2^{+}}
\newcommand{\formo}{\mathrm{C}_N \mathrm{H}_2^{\,0}}

\newcommand{\cfourty}{\mathrm{C}_{40} \mathrm{H}_2}

\newcommand{\ceighty}{\mathrm{C}_{80} \mathrm{H}_2}
\newcommand{\chundred}{\mathrm{C}_{100} \mathrm{H}_2}
\newcommand{\Etot}{E_{\mathrm{tot}}}
\newcommand{\EA}{\mathrm{EA}}
\newcommand{\IP}{\mathrm{IP}}
\newcommand{\Elum}{E_{\mathrm{LUMO}}^0}
\newcommand{\Ehom}{E_{\mathrm{HOMO}}^0}
\newcommand{\Upol}{U_{\mathrm{pol}}}
\newcommand{\dn}{\delta_n}

\usepackage{graphicx,amssymb,amsmath}

\begin{document}

\title{Model \textit{ab initio} study of charge carrier solvation and large polaron formation on conjugated carbon chains}

\author{M.~L.~Mayo}
\affiliation{Department of Physics, The University of Texas at
Dallas, P. O. Box 830688, EC36, Richardson, Texas 75083, USA}
\author{Yu.~N.~Gartstein}
\affiliation{Department of Physics, The University of Texas at
Dallas, P. O. Box 830688, EC36, Richardson, Texas 75083, USA}

\begin{abstract}
Using long $\form$ conjugated carbon chains with the polyynic structure as  prototypical examples of one-dimensional (1D) semiconductors, we discuss self-localization of excess charge carriers into 1D large polarons in the presence of the interaction with a surrounding polar solvent. The solvation mechanism of self-trapping is different from the polaron formation due to coupling with bond-length modulations of the underlying atomic lattice well-known in conjugated polymers. Model \textit{ab initio} computations employing the hybrid B3LYP density functional in conjunction with the polarizable continuum model are carried out demonstrating the formation of both electron- and hole-polarons. Polarons can emerge entirely due to solvation but even larger degrees of charge localization occur when accompanied by atomic displacements.
\end{abstract}


\maketitle
\section{\label{Intro}Introduction}

One-dimensional (1D) semiconductor (SC) structures such as $\pi-$conjugated polymers, nanotubes and nanowires are interesting nanoscopic objects that can be exploited in various areas, including (opto)electronics, energy harvesting and sensors. The nature and properties of excess charge carriers on these structures are fundamental for many applications. As covalent bonds provide for wide electronic bands and effective band masses much smaller than the free electron mass, intrinsically excess charge carriers in these systems have a high propensity for delocalization and a potential for high mobilities. If not for extrinsic defects, inherent limitations on the mobility arise due to interactions of charge carriers with other subsystems -- most commonly with displacements of the underlying atomic lattice of the 1D SC structure. This electron-phonon coupling has been a subject of numerous studies for various specific 1D systems. It is well-known that, as a result of the interaction with lattice phonons,  excess charge carriers may undergo self-localization into polaronic states. Such polarons have been extensively discussed in the context of conjugated polymers (CPs), where they are believed to be accompanied by localized bond-length modulation patterns and new features in the optical absorption (see, e.g., Refs.~\cite{YuLubook,Bredas1985} for reviews and original references).

A different and much less explored implementation of the strong polaronic effect can take place when a 1D semiconductor is immersed in a 3D polar medium, the situation of particular relevance to applications and processes involving fundamental redox reactions in polar solvents. In this case of excess charge carrier \textit{solvation}, the long-range Coulomb interaction, as has been recently discussed  within simplified theoretical models \cite{conwell_1,YNGpol,polcylinder,GU_lowfreq,UG_optabs}, can result in the formation of 1D adiabatic large-radius polarons, where a localized electronic state on a SC structure is surrounded by a self-consistent pattern of the sluggish (orientational) polarization of the solvent. The Coulomb mechanism of the polaron formation here is analogous to the well-known 3D polarons in polar SCs \cite{polarons1,appel,polarons2,CTbook} and solvated electrons in polar liquids \cite{CTbook,ferra91,nitzan}.  Distinct from those cases are the confinement of the electron motion in a 1D nanostructure and its structural separation from the 3D polarizable medium. The energetic significance of the solvation has been emphasized by finding \cite{YNGpol,polcylinder} that the binding energy of the resulting 1D polarons could reach a substantial fraction, roughly one third, of the binding energy of Wannier-Mott excitons, the well-known primary photoexcitations in many 1D SCs. This may lead to enhanced charge separation. On the other hand, the mobility of solvated charge carriers is drastically reduced due to the dissipative drag of the medium \cite{conwell_1,CBDrag,GU_lowfreq}. New optical absorption signatures are also expected as a result of solvation \cite{UG_optabs,MGshort_prb}.

Given the generic character of the solvation mechanism of charge-carrier self-localization and its potential implications for carrier mobilities and redox processes on 1D SC nanostructures, this effect appears interesting and important enough to warrant studies at various computational levels. As a step in that direction, we have attempted a first -- to our knowledge -- model \textit{ab initio} study reported in a recent communication \cite{MGshort_prb} that would demonstrate the formation of \textit{solvation-induced} 1D large-radius polarons. In this paper we elaborate on that brief report discussing features of solvation-induced self-localization of excess charges on finite linear conjugated carbon chains $\form$ (polyyne oligomers) as derived within density-functional theory (DFT) computations in conjunction with the polarizable continuum model (PCM) \cite{g03,leach_1,QMCSM}.

As in numerous studies of electron-lattice polarons in CPs, \textit{ab initio} studies are expected to be helpful in elucidating the role played by valence electrons and many-electron interactions in accommodation of an extra charge carrier. It should be noted that a reorganization of valence band electrons can be important even in non-interacting electron models -- a nice illustration was given in Ref.~\cite{FVB} that analyzed how polarons of a two-band Peierls dielectric model evolve into single-particle Holstein \cite{holstein_LP} polarons in the limit of the ``frozen valence band'' approximation.  With realistic Coulomb interactions in place, the role of valence electrons in the formation of the relevant self-consistent potentials would only increase. Interestingly enough, despite a relatively long history of first-principles studies of electron-phonon polarons in CPs (see, e.g., multiple references in  Refs.~\cite{BredasPolOptics,bobbert}), certain questions have been raised recently regarding applicability of different \textit{ab initio} frameworks to describe the formation of those polarons. That pertains to the failure of the local-density-approximation and generalized-gradient-approximation DFT schemes to detect self-localized charge density distributions in various charged oligomers, while Hartee-Fock, parameterized semi-empirical and possibly hybrid-functional DFT methods have been reported to lead to charge localization in the middle of oligomers (e.g., Refs.~\cite{bobbert,pitea_1,bredas_3,zuppi2003}). Quite different level-of-theory-dependent magnitudes of the effective electron-phonon coupling have also been found even in neutral polymeric systems (see, e.g., a comparative discussion in Refs.~\cite{yang_kertesz_1,yang_kertesz_2}). A caution therefore has to be exercised in interpretation of at least the quantitative aspects of various \textit{ab initio} results.

As first-principles calculations are applied to the effect of solvation on charge-carrier self-localization within the PCM framework, it would also be desirable to have a comprehensive comparison of different  computational approaches. Such a study is currently underway and we are planning to report its results elsewhere. Reassuringly, the effect of solvation on excess charge localization appears qualitatively robust in those computations. While providing an illustrative example of comparison with  Hartree-Fock computations, in this paper we rather focus on a more modest goal to discuss trends that are derived within a single computational scheme. Specifically, we employ DFT calculations with the B3LYP hybrid exchange-correlation density functional \cite{Martin}. This functional is known to include both local and non-local effects and has been commonly used in recent studies \cite{yang_kertesz_1,yang_kertesz_2,schaefer_1,schaefer_2} of polyyne and its oligomers.

By exploring model systems with prescribed carbon atom positions, we will be able to illustrate excess charge carrier (an electron or a hole) localization and the resulting formation of large 1D polarons entirely due to the polarization of the surrounding solvent. Examples of self-consistent -- that is, practically independent of the chain length -- polaron structures will thereby be also given. Within the confines of the computational scheme used, ``realistic'' geometries of carbon atom arrangements are those that optimize the total energy of the system. We have performed such geometrical optimizations for polyynic chains both in vacuum and in solvent. In the range of chain lengths studied (up to $N=100$ of carbon atoms) with the B3LYP hybrid functional, we have not been able to detect a clear trend towards excess charge self-localization due to the electron-phonon interaction \textit{alone} -- as the chain length increased, the bond-length modulation pattern would only become flatter in the middle of the chain. On the contrary, in the presence of the Coulomb interaction with the solvent, clearly localized patterns of bond-length alternation have been found to act as to boost the degree of excess charge localization, possibly reflecting synergistic effects from both mechanisms in this case.

\section{\label{Comps}Systems, Computations, and Data Analysis}

Since the computational demand increases substantially in DFT-PCM calculations,  we have selected structurally simple 1D semiconducting systems of hydrogen-terminated polyynic linear carbon chains $\form$  for our demonstration. It should be stressed however that, while playing the role of a prototypical example in our study, polyynic chains continue to be the subject of much attention in their own right (\cite{gladysz_1,gladysz_2,yang_kertesz_1,yang_kertesz_2} and multiple references therein). Interestingly, they were predicted \cite{rice_2,rice_3} to possess a rich family of electron-lattice self-localized excitations because of the extra degeneracy of molecular orbital (MO) levels.

The subject of our attention in this paper are long chains $\form$ with even number of carbons $N$ ($N$ between 20 and 100). It is well established now  \cite{karpfen_1,yang_kertesz_1,yang_kertesz_2,pfeiffer_1,schaefer_1} that even-$N$ long ($N \gtrsim 10$) neutral chains  have their ground state with the polyynic structure, that is, they feature a (nearly uniform) alternating pattern of triple $\triple$ and single $\single$ bonds leading to a gap in the electronic spectrum. The ground state of odd-$N$ chains would exhibit a kink-like ``defect'' in the bond-alternation pattern with associated mid-gap states \cite{rice_2,schaefer_2}, which we do not discuss here. Having neutral chains $\formo$ as a benchmark, we have studied both negatively charged $\formmin$ (an extra electron) and positively charged $\formpl$ (an extra hole) chains. As is clearly illustrated below, their self-localization behavior has been found very similar, with the  excess charge measured correspondingly in units of $(-e)$ or $(e)$, where $e$ is the magnitude of the fundamental charge. To save space,  some of the results, therefore, would be shown only for the negatively charged systems.

\begin{figure*}
\centering
\includegraphics[scale=0.83]{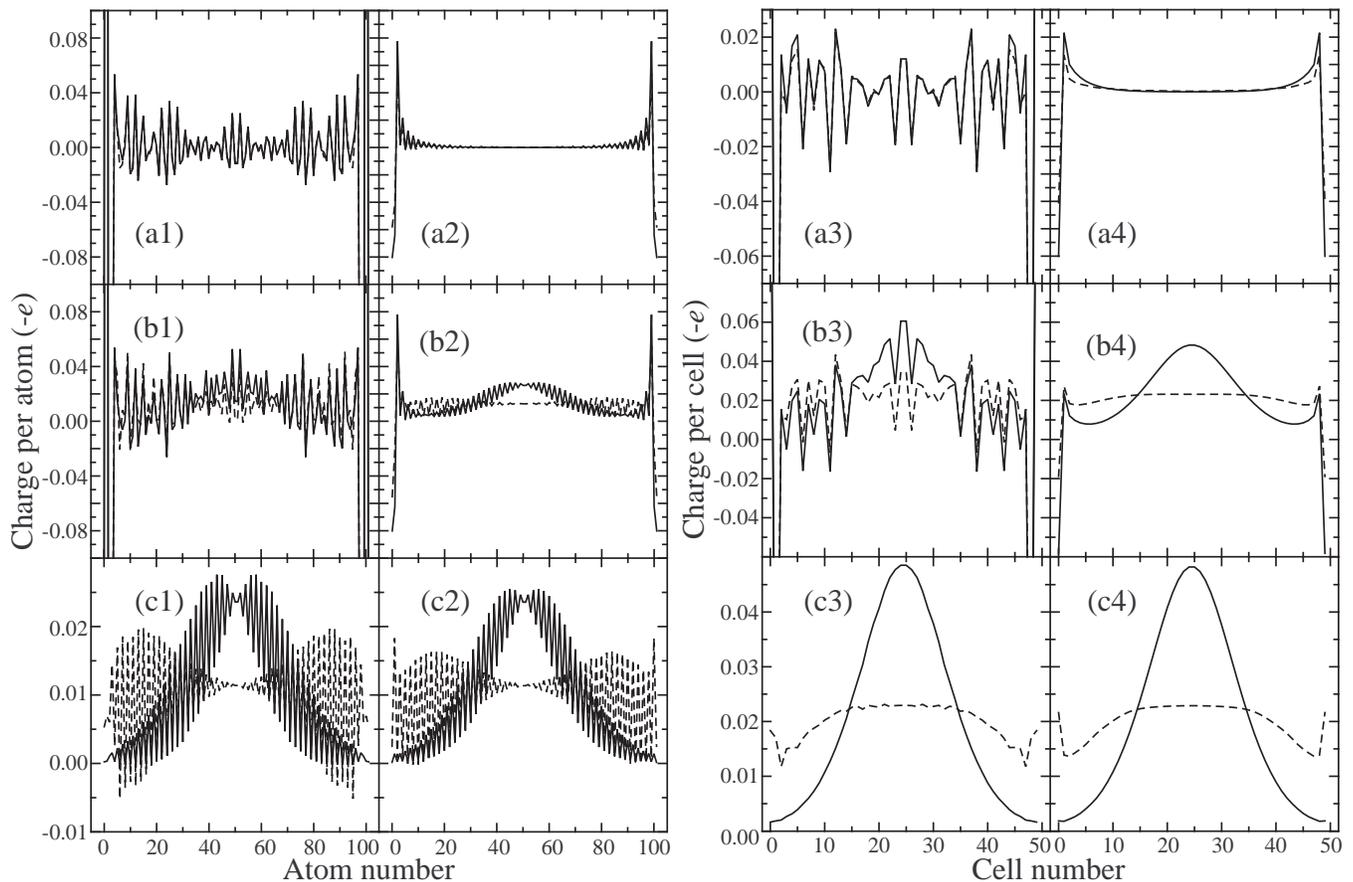}
\caption{\label{MulVsLow}Spatial charge distributions on $N=100$ polyynic chains with RG1 geometry derived by using atomic charges from Mulliken (panel columns indexed 1 and 3) and L\"{o}wdin (panel columns 2 and 4) procedures. Panel row (a) is for the totally neutral chains, row (b) for the negatively charged chains with one added electron; row (c) depicts the distribution of \textit{excess} charge defined as a difference of the data in rows (b) and (a). Two left columns (1 and 2) depict charges per atom, two right columns (3 and 4) yield charges per unit cell. In each of the panels, solid lines show the results for chains in solvent, whereas dashed lines (when distinguishable) for chains in vacuum. It is evident that cell-centric excess charge distributions in panels (c3) and (c4) are nearly identical despite substantial differences between Mulliken and L\"{o}wdin charges observed in other panel pairs.}
\end{figure*}

For the \textit{ab initio} engine in this study we employ density functional theory (DFT) within the Gaussian 03 \cite{g03} set of programs. More specifically, the majority of the results presented in this paper have been derived with the B3LYP 
hybrid exchange-correlation functional which includes the Hartree-Fock exchange along with DFT correlation \cite{Martin}.
In order to investigate the effects of solvation in a polar medium, Gaussian 03 \cite{g03} offers its implementation of the Polarizable Continuum Model (PCM) described in original publications \cite{tomasi_1,tomasi_2,tomasi_3,barone_1}. A rich 6-311++G (\textit{d,p}) all electron basis set has been employed throughout. It should be noted that the results of our ``in vacuum'' calculations have been verified to compare well against previously published \textit{ab initio} data on neutral \cite{yang_kertesz_2} and charged \cite{schaefer_1} polyynic systems both in terms of energetics and optimized bond-lengths. For all DFT-PCM ("in solvent") calculations, water has been chosen as a polar solvent with its default parameters in Gaussian 03. We remind the reader that the focus of this paper is to compare results obtained in vacuum and in solvent environments within the same computational scheme rather than compare various computational approaches.

In order to be able to clearly discern the effects due to the solvation and provide conceptual proofs and illustrations (see Sec.~\ref{Models}), we will be exploring both systems with the full geometric optimization of the underlying atomic lattice and systems with \textit{model} rigid geometries, that is, with prescribed atomic positions. One of such rigid geometries, referred to as RG1, features the following fixed lengths: 1.235~\AA \ for the triple bonds, 1.326~\AA \ for the single, and 1.063~\AA \ for the $\ch$ bonds. These lengths have been retrieved from our complete optimization of the neutral $N=80$ chain in vacuum and are in fact relatively close to optimal values (see Sec.~\ref{BLA}). Other rigid geometries have also been explored where all single $\single$ bonds are artificially stretched in successive increments of 0.1~\AA \ up to the final length of 1.726~\AA, the latter geometry is denoted as RG2. This artificial stretching is used to study the effect of narrower electronic bandwidths (larger band effective masses) on the charge localization.

A very relevant quantity for our discussion is the equilibrium \textit{spatial charge distribution} over the 1D SC, which is naturally related to atomic charges in outputs of \textit{ab initio} calculations. Importantly, these charges are calculated from many-electron wave functions and thereby reflect responses due to \textit{all} electrons in the system. Two procedures, due to Mulliken and L\"{o}wdin, are widely used for charge population analysis \cite{leach_1}. It is well known \cite{leach_1} that calculations of atomic charges depend on the basis set transformations and sometimes lead to artifacts and spurious results. Such artifacts are in fact apparent in our illustration in Fig.~\ref{MulVsLow} that compares various results we obtained for $N=100$ chains. Differences between Mulliken and L\"{o}wdin charges are especially clearly seen for raw charge populations separately on neutral (row (a)) and charged (row (b)) chains, whether they are atomic- or unit-cell-centric. Comparison of panels (c1) and (c2), however, indicates that certain cancelation effects take place and both methods lead to less dissimilar results when one is interested in the spatial distribution of the \textit{excess} charge derived as the difference of charge densities on charged and neutral chains. Still those atomic-centric results exhibit spurious oscillations which may be artifacts related to the presence of bond-oriented charge-density waves in our system. This suggests ``averaging'' of the charge over the unit cells consisting of pairs of neighboring carbons (for convenience, the chain end cells are defined here to include the hydrogen atomic charges). Panels (c3) and (c4) show that various artifacts indeed practically disappear, and the unit-cell-based excess charge distributions calculated by Mulliken and L\"{o}wdin procedures become remarkably close. We confirmed this conclusion for many other cases of interest. Accordingly, it is these \text{stable} results for excess charge per cell that will be displayed in our results below (using raw L\"{o}wdin charges).

\section{\label{Models}Solvation-induced self-consistent charge localization in model systems}

In real systems, both atomic displacements and polarization of the surrounding medium are expected to take place in order to accommodate an excess charge carrier. Moreover, given the nonlinear nature of the polaronic effect, both subsystems may act in a synergistic way. In order to better understand the contribution coming from the solvation, we therefore start from studying model systems with rigid geometries, where no localized bond-alternation patterns are allowed while the medium ``adjusts'' its state of polarization.

\subsection{\label{ExcCharge}Spatial distribution of the excess charge}

\begin{figure}
\centering
\includegraphics[scale=1.0]{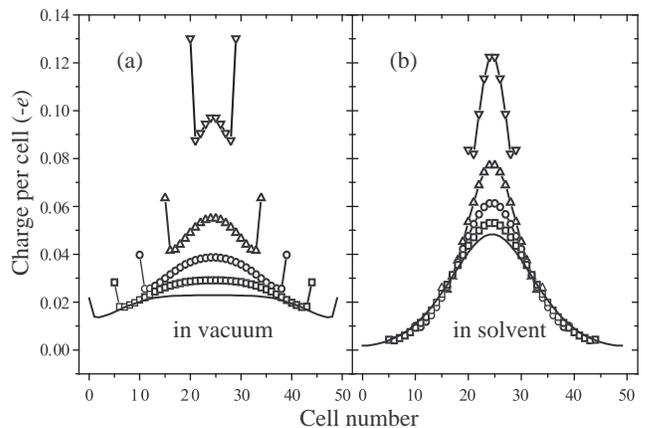}
\caption{Excess charge density spatial distribution on negatively charged $\formmin$ chains with  $N=20$, 40, 60, 80 and 100 -- top to bottom curves, respectively, in their central parts. All chains feature the same pattern of fixed bond-lengths corresponding to RG1 geometry described in text. Panel (a) displays results for chains in vacuum, panel (b) for chains in the solvent environment.}
\label{fig:ch_diff_r0}
\end{figure}

\begin{figure}
\centering
\includegraphics[scale=0.83]{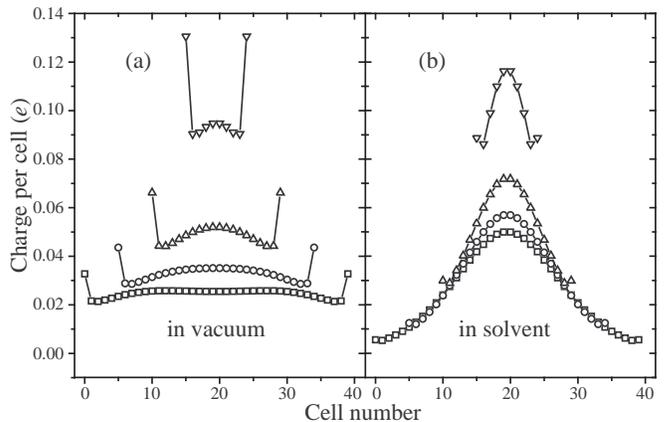}
\caption{As in Fig.~\ref{fig:ch_diff_r0} but for positively charged $\formpl$ chains with  $N=20$, 40, 60 and 80 -- top to bottom curves, respectively, in their central parts.}
\label{fig:ch_diff_r0_plus}
\end{figure}

Figures \ref{fig:ch_diff_r0} and \ref{fig:ch_diff_r0_plus} compare excess charge density distributions, negative and positive, respectively, calculated as described in Sec.~\ref{Comps}, for chains of varying lengths both in vacuum and in solvent.
It is evident that panels (a) and (b) of each of the figures display quite distinct trends. Disregarding the end effects (end cells include charges on hydrogens), vacuum results in panels (a) reflect the ``particle in a box'' behavior: as the chain length increases, the excess charge is distributed more and more uniformly over the whole chain. In a sharp contrast, for chains in the solvent, panels (b), the excess charge exhibits much more localized distributions around central parts of the chains, to the extent that no appreciable end effects are present. As the chain length increases, those distributions show a very clear tendency towards convergence, albeit not completely achieved within the range of lengths studied.

\begin{figure}
\centering
\includegraphics[scale=1.0]{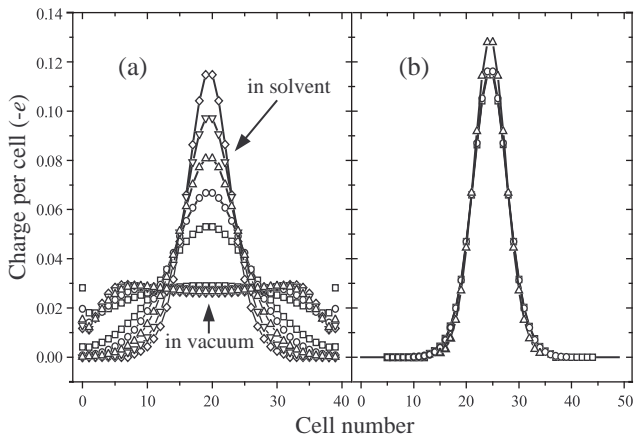}
\caption{Excess charge density spatial distribution on charged $\formmin$ chains. (a) N=80. The evolution of the charge distribution upon stretching the single bonds by 0.1 {\AA} \ successive increments from 1.326 {\AA} \ to 1.726 {\AA} \ -- bottom to top in a group of 5 clearly distinct curves shown by different connected symbols and denoted ``in solvent''. The same - but unconnected - data symbols are used to show the results for chains in vacuum, those distributions largely overlap. (b) Comparison of distributions for chains in the solvent with $N=40$, 60, 80 (different data symbols) and 100 (solid line) for the geometry RG2 featuring the longest single bonds. Curves for $N=60$, 80 and 100 practically coincide.}
\label{fig:combined_minus}
\end{figure}
\begin{figure}
\centering
\includegraphics[scale=0.8]{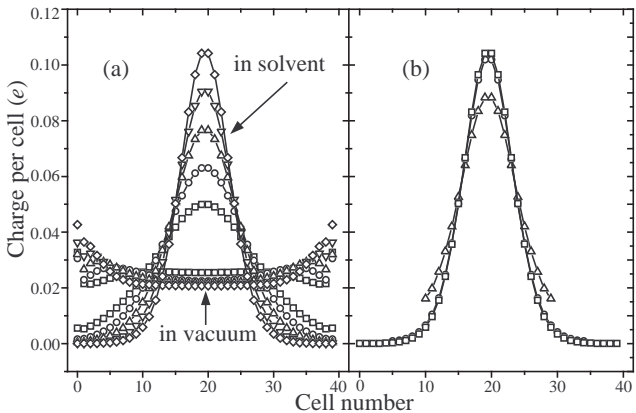}
\caption{Excess charge density spatial distributions on charged~$\formpl$ chains. (a) N=80.
The evolution of the charge density distribution upon stretching bonds as in Fig.~\ref{fig:combined_minus}. (b) Comparison
of distributions for chains in the solvent, with~$N=40$, 60, 80 (different data symbols) for the RG2 geometry. Curves for $N=60$ and 80 practically coincide.}
\label{fig:combined_plus}
\end{figure}

Instead of further increasing the chain length in already demanding computations, we now attempt to decrease the spatial extent of charge localization in model systems. To this end, we recall that, in single-particle models \cite{conwell_1,YNGpol,polcylinder,GU_lowfreq,UG_optabs}, the spatial extent of the polaron  is determined by the balance of the gain in the potential energy of the self-localized carrier due to its interaction with the medium, and the loss in its kinetic energy due to localization. Decreasing the role of the kinetic energy would lead to a diminished delocalization propensity and is expected to shorten the polaron. This should be achievable with an increased effective band mass of the carrier, or with a narrower electron bandwidth. To affect this, we chose to \textit{artificially} stretch all single bonds in our rigid model system as described in Sec.~\ref{Comps}, thereby making $\triple$ ``dimers'' more and more separated from each other. Figures \ref{fig:combined_minus}(a) (extra electron) and \ref{fig:combined_plus}(a) (extra hole) illustrate a \textit{dramatic} difference in the response of the excess charge distributions on $N=80$ chains to successive single-bond stretchings in vacuum and in solvent. Whereas the charge density in vacuum shows only very little changes in this process and preserves nearly uniform distributions over the whole chain, for chains in solvent, each successive stretching step indeed results in substantially smaller and smaller localization regions. As the extent of localization in the longest-stretched geometry, RG2, appears to be well within the chain length, we can now compare the corresponding excess charge distributions for several chain lengths. Figure \ref{fig:combined_minus}(b) makes it evident that the distributions for $N=60$, 80, and 100 negatively charged chains in solvent practically coincide. Likewise, distributions for $N=60$ and 80 positively charged chains are practically identical in Fig.~\ref{fig:combined_plus}(b). This confirms that the obtained charge distributions on longer chains correspond to a fully converged self-consistent pattern of excess charge localization. Importantly, self-localization of the excess charge demonstrated here has been achieved \textit{entirely} due to the interaction with the surrounding solvent.

Comparing results for negatively charged and positively charged species, Fig.~\ref{fig:ch_diff_r0} vs Fig.~\ref{fig:ch_diff_r0_plus} and Fig.~\ref{fig:combined_minus} vs Fig.~\ref{fig:combined_plus}, one, of course, notices some effects of charge-conjugation (CC) symmetry breaking. They depend on the chain length as affected by both the ends and by the inherent band structure. As the inherent $\pi$-orbital bands are nearly CC-symmetric, these effects are evidently relatively weak, and basically the localization behavior for electrons and holes appears very similar.

\subsection{Molecular orbitals}

\begin{figure}
\centering
\includegraphics[scale=0.73]{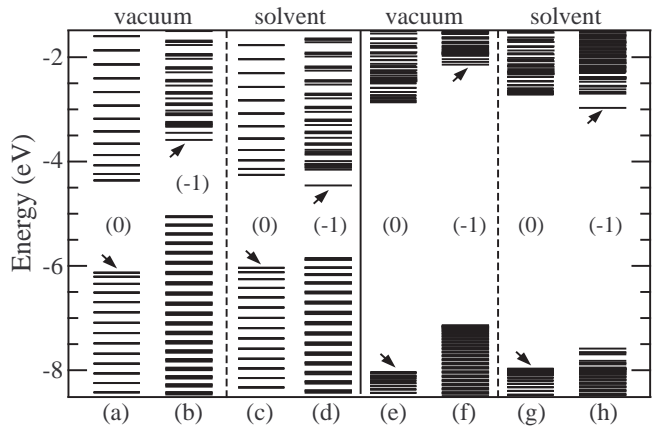}
\caption{Molecular orbital energy levels for neutral (0) and charged (-1) C$_{100}$H$_2$ chains with rigid RG1 (a-d) and RG2 (e-h) geometries both in vacuum and in the solvent. The arrows indicate positions of HOMO levels.}
\label{fig:v6-c100h2_bands_r0_r4}
\end{figure}

\begin{figure}
\centering
\includegraphics[scale=0.73]{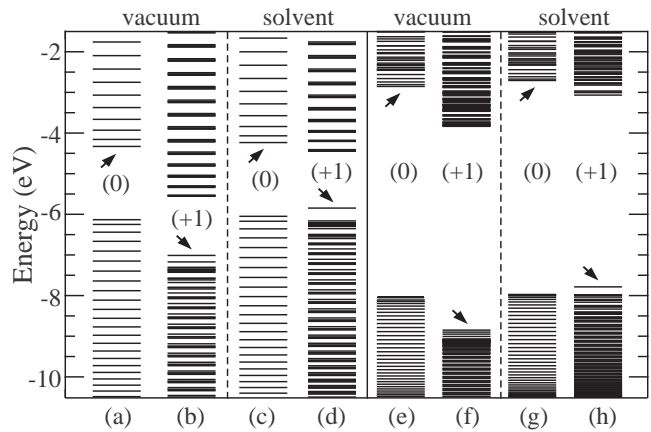}
\caption{Molecular orbital energy levels for neutral (0) and charged (+1) C$_{80}$H$_2$ chains with rigid RG1 (a-d) and RG2 (e-h) geometries both in vacuum and in the solvent. The arrows indicate positions of LUMO levels.}
\label{fig:v1-c80h2_plus1_bands}
\end{figure}

\begin{figure*}
\centering
\includegraphics[scale=.32]{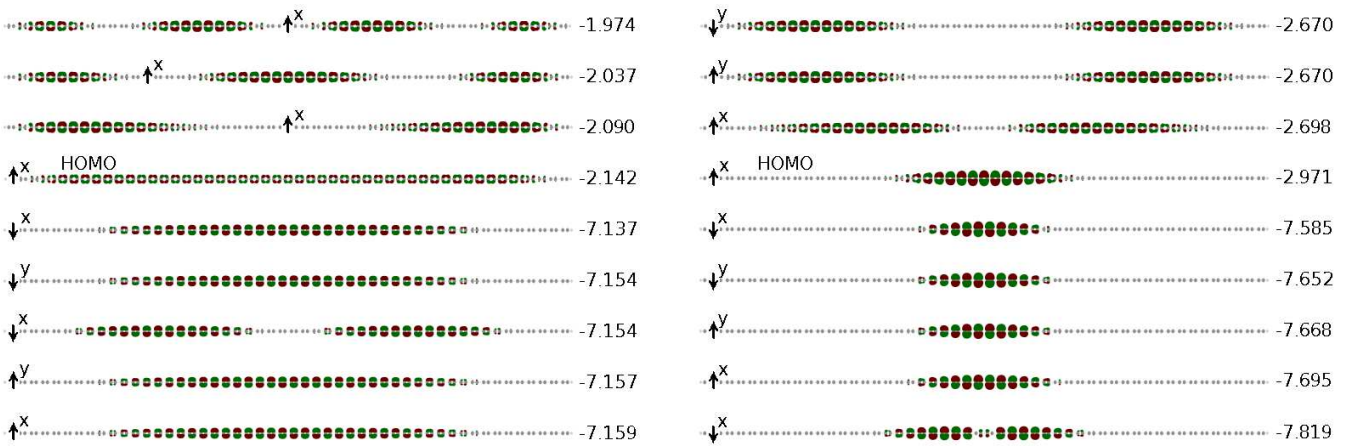}
\caption{Spatial distribution of several MOs for $\chundred^{-}$ chains with RG2 geometry in vacuum (left column) and in solvent (right column).   The spin state of orbitals is indicated by arrows, while the spatial orientation by letters ``x'' and ``y'' (chains are oriented along the z-axis). Orbital energies in eV are shown to the right of the columns. Compared MOs around HOMOs are sequentially selected from a list ordered in orbital energies.}
\label{fig:MO_minus1_combined}
\end{figure*}

We now turn from the total excess charge distribution to individual MOs and the resulting electronic structure. In the neutral $\form$ chains, $\pi$-electron energy levels exhibit a four-fold degeneracy -- in addition to the spin degeneracy, there is a two-fold degeneracy with respect to the spatial orientation of  $\pi$-orbitals. If the chain direction is chosen as the $z$-axis, $\pi$-orbitals can be said to be oriented either along $x$ or along $y$ axes. In charged chains, spin- and orbital-orientation-degeneracy are generally lifted. Also, in the presence of the polaronic effect, one observes both spatially localized and delocalized (over the chain length) individual electronic states. The overall picture of MO energy levels in neutral and charged chains is illustrated in Figs.~\ref{fig:v6-c100h2_bands_r0_r4} ($\chundred^{0}$ vs $\chundred^{-}$) and \ref{fig:v1-c80h2_plus1_bands} ($\ceighty^{0}$ vs $\ceighty^{+}$), while Fig.~\ref{fig:MO_minus1_combined} illustrates the spatial behavior of MOs on charged $\chundred^{-}$ chains in more detail.

Looking at the electronic structures of neutral and charged chains in Figs.~\ref{fig:v6-c100h2_bands_r0_r4} and \ref{fig:v1-c80h2_plus1_bands}, one can compare the differences in responses between chains in vacuum and chains in solvent as well as compare the differences caused by the geometry of carbon atom arrangements (RG1 vs RG2).  Going to the artificial RG2 geometry with longer $\single$ bonds allows us here to decrease the polaron size and to illustrate some features that would otherwise be hard to observe within the range of chain lengths studied. Probably the first visually noticeable difference between RG1 and RG2 electronic structures is a much wider band gap in RG2 -- as indeed one should expect based on a much larger difference in the hopping integrals corresponding to triple and single bonds. As such, however, the size of the band gap is inconsequential for the degree of charge localization due to solvation. Instead, of direct importance here in a much narrower bandwidth in RG2, which is easily detected via a higher density of states (panels (e) vs (a) and (g) vs (c)). The narrower the bandwidth is, the smaller is the propensity of the excess carrier to delocalize.

A common feature evident in results for charged chains in vacuum is appreciable overall energy shifts with respect to neutral chain MO levels: to higher energies for excess electrons and to lower energies for excess holes. These shifts reflect simple electrostatic effects (the electrostatic potential with respect to infinity) for chains of finite lengths.  Understandably, shifts of such an origin should become much smaller for chains surrounded by media with high dielectric constants, which is indeed the case for our chains in the solvent environment (see also below in Sec.~\ref{Energetics}). A much more important and relevant difference in the MO structure of charged chains in vacuum and in solvent is instead in that solvated chains feature local electronic levels clearly separated from other states. In the case of negatively charged chains, Fig.~\ref{fig:v6-c100h2_bands_r0_r4}, it is the highest occupied molecular orbital (HOMO) accommodating an excess electron. In the case of positively charged chains, Fig.~\ref{fig:v1-c80h2_plus1_bands}, it is the lowest unoccupied molecular orbital (LUMO) accommodating an excess hole. Comparison of panels (d) to (b) and, especially, panels (h) to (f) in these figures clearly demonstrates the formation of local levels - as would be caused by the appropriate polarization of the environment. The appearance of localized electronic states in self-consistent ``potential wells''  is one of the hallmarks of the strong polaronic effect \cite{appel,CTbook}.

The number of localized electronic states appearing due to the polaronic effect should depend on the nature of the system and the effective potential well. For electron-lattice polarons in CPs, local levels are known \cite{YuLubook,Bredas1985} to split off from both conduction and valence bands. The short-range character of the effective potential well in that case, however, limits the number of such states (in traditional two-band Peierls dielectric models \cite{YuLubook}, there would be only two local levels: one close to the conduction and one to the valence band). In the case of the long-range interaction with the polarizable environment, the effective potential well has the long-range Coulomb behavior expected to result, for an infinite system, in a multitude of localized states of various spatial symmetries converging to the onset of the delocalized continuum. This is a standard picture for traditional ``single-band'' 3D polarons \cite{appel}, and we have specifically discussed its realization for 1D systems immersed in the 3D polar medium \cite{YNGpol,UG_optabs}.

Restricting a more detailed illustration now to the negatively charged $\chundred^{-}$ chains, Fig.~\ref{fig:MO_minus1_combined} compares the spatial structure of HOMOs and several other MOs (from a list ordered in orbital energies) in vacuum and in the solvent environment as retrieved from the GaussView 4 \cite{gv41} rendition of Gaussian 03 outputs. Here we use results only for RG2 geometry displaying better distinguishable localized states. (The figure of course also illustrates the lifting of spin and spatial-orientation degeneracies.) As expected, a stark difference is seen between the HOMOs in vacuum and solvent cases: the former shows a state delocalized over the chain length, while the latter indeed exhibits a well-localized character. In addition to this, however, we also find in the right column of Fig.~\ref{fig:MO_minus1_combined} that the solvation-induced negative polaron features various occupied localized states in the valence band region of energies (nothing like this happens for valence band states in the vacuum case). This observation confirms a reorganization of valence band states apparent in the corresponding electronic structure of Fig.\ref{fig:v6-c100h2_bands_r0_r4}(h). Thus the polaron reveals a many-electron structure, more complex than invoked in simplified single-particle models \cite{conwell_1,YNGpol,polcylinder,GU_lowfreq,UG_optabs}.

Even in RG2 geometry it is difficult though to visually identify the LUMO in the right column of Fig.~\ref{fig:MO_minus1_combined} as an unoccupied localized state expected to be formed in the effective polarization potential well \cite{YNGpol,UG_optabs}. While that state wave function evidently exhibits the right odd symmetry with respect to the inversion of the chain axis, the length of the chain appears insufficient to establish a localized character of the state with certainty. Likewise, within the length of our oligomer we cannot establish the onset of what would become a continuum of truly delocalized states in the limit of infinite chains.

Similar considerations of details of the spatial behavior of MOs can be done for positively charged chains which would then correlate with signatures of the electronic structure in Fig.~\ref{fig:v1-c80h2_plus1_bands}.

\subsection{\label{Energetics}Energetics}

An important energetic quantity in discussions of the polaronic effect is the polaron binding energy referring essentially to the gain in the \textit{total} energy of the system taking place upon self-localization of the excess charge carrier \cite{YuLubook,appel,CTbook}. Calculations of the binding energy of solvation-induced polarons on 1D structures are very straightforward in simplistic models \cite{conwell_1,YNGpol,polcylinder,GU_lowfreq,UG_optabs}. Unfortunately, we do not feel we have a reliable straightforward procedure for determination of the binding energy of solvation-induced polarons in our current \textit{ab initio} calculations. One of the reasons for this is that we cannot control the state of the polarization of the solvent as it would be easily done by prescribing desired patterns of atomic displacements in the case of electron-lattice polarons. Some considerations can, however, be put forward based on direct outputs of Gaussian 03 computations.

\begin{figure}
\centering
\includegraphics[scale=0.52]{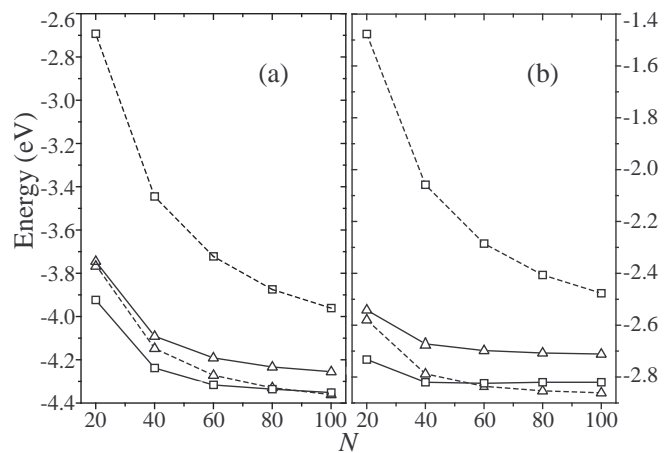}
\caption{The negative of electron affinities, $-\EA$, (square symbols) and energies $\Elum$ of neutral chains (triangles) as functions of the number of carbons in polyyne oligomers $\form$ with rigid geometries. Vacuum data points are connected by dashed lines,  data for chains in solvent by solid lines. Panel (a) displays results for chains with RG1 geometry, panel (b) corresponds to chains with RG2 geometry.}
\label{fig:ea-lumo}
\end{figure}

Figure \ref{fig:ea-lumo} illustrates such considerations for excess electrons. One very well-defined and physically meaningful quantity characterizing accommodation of an excess electron in the computational outputs is the electron affinity
$$
\EA=\Etot^{0}-\Etot^{-},
$$
corresponding to the difference of total system energies for neutral (superscript ``0'') and negatively charged \mbox{(``-'')} chains.  Another readily available energy is the energy $\Elum$ of the LUMO in the neutral system. These energies are compared in Fig.~\ref{fig:ea-lumo} as $-\EA$ vs $\Elum$  for various conditions as functions of the number $N$ of carbons in $\form$. The rationale for this comparison is that in the limit of infinite chains, $N\rightarrow \infty$, one might hope, in the spirit of Koopman's theorem \cite{Martin,Szabo}, that $\Elum$ would represent the extra system energy upon an addition of a single electron into the lowest-energy delocalized state with vanishing excess charge density. The negative of the electron affinity, $-\EA$, on the other hand, represents the extra system energy upon addition of a single electron when the system is allowed to completely reorganize itself to achieve the minimum possible total energy. Both compared ``extra energies'' are of course negative in our data. We remind the reader that in this section we discuss system with rigidly positioned carbons so that reorganization here is limited only to the states of other electrons and the polarization of the environment.

A striking difference is observed in Fig.~\ref{fig:ea-lumo} when results for chains in vacuum are compared to chains in solvent. In the former case the data corresponds to $-\EA > \Elum$, while in the latter the order of these energies is reversed: $-\EA < \Elum$, clearly showing the significance of screening and reorganization effects taking place in the solvent environment. Vacuum data points are evidently very far from displaying a convergence to the infinite chain limit, and a larger steepness of the vacuum $\EA$ as a function of $N$ is indicative of the magnitude of the Coulomb effects involved for finite excess charge densities in oligomers. While data points for chains in solvent, of course, also exhibit an $N$-dependence, this dependence is shallower than in the vacuum case, and for RG2 geometry (panel (b)) displays a nearly converged behavior for largest $N$ studied. As we discussed in Sec.\ref{ExcCharge}, a self-consistent polaron formation is achieved in RG2 within our range of chain lengths.

Considering the positive energetic ``improvement'' $\Elum + \EA$ in the solvent environment, its $N=100$ magnitudes in Fig.~\ref{fig:ea-lumo} are equal to approximately 0.1 and 0.11 eV for RG1 and RG2 geometries, respectively.
If one were to take  this energetic improvement in the infinite-$N$ limit  as corresponding to the ``true'' solvation-induced polaron binding energy, then the obtained numbers $\simeq 0.1$ eV would provide estimates of the polaron binding.

\begin{figure}
\centering
\includegraphics[scale=0.52]{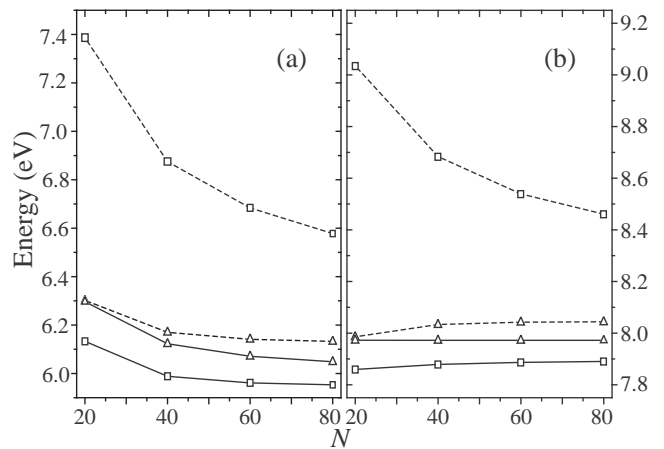}
\caption{Ionization potentials, $\IP$, (square symbols) and the negative of HOMO energies, $-\Ehom$, of neutral chains (triangles) as functions of the number of carbons in $\form$ with rigid geometries. Vacuum data points are connected by dashed lines,  data for chains in solvent by solid lines. Panel (a) displays results for chains with RG1 geometry, panel (b) corresponds to chains with RG2 geometry.}
\label{fig:ip-homo}
\end{figure}

An analogous consideration for excess holes involves a comparison of the ionization potential
$$
\IP = \Etot^{+} - \Etot^{0}
$$
and the negative $-\Ehom$ of the energy of the HOMO in the neutral chain. This comparison of now positive extra system energies is illustrated in Fig.~\ref{fig:ip-homo}. The figure displays qualitative patterns similar to observed above for excess electrons (while, of course, also showing signs of CCS-breaking). Particularly, one notes while for chains in vacuum one has $\IP > -\Ehom$,  the relationship reverses for chains in solvent: $\IP < -\Ehom$. The $N=80$ values of the energetic ``improvement'' $-\Ehom-\IP$ in the solvent environment are approximately 0.1 and 0.08 eV for RG1 and RG2 geometries, respectively, which could again suggest polaron bindings close to 0.1 eV.

It appears, however, strange that these naive estimates lead to relatively small differences (and even to a reversal) in polaron bindings for RG1 and RG2 geometries -- while the spatial localization has been found much more significant in RG2, Sec.~\ref{ExcCharge}. We believe this indicates that the actual polaron binding energies would be underestimated in discussed energetic comparisons. It is quite likely that $\Elum$ and $-\Ehom$ do not represent fully the lowest extra energy of systems  where an excess delocalized electron or an excess delocalized hole would interact with a non-uniform, charge density wave (CDW), background  of the polymer chain.

\begin{figure}
\centering
\includegraphics[scale=0.9]{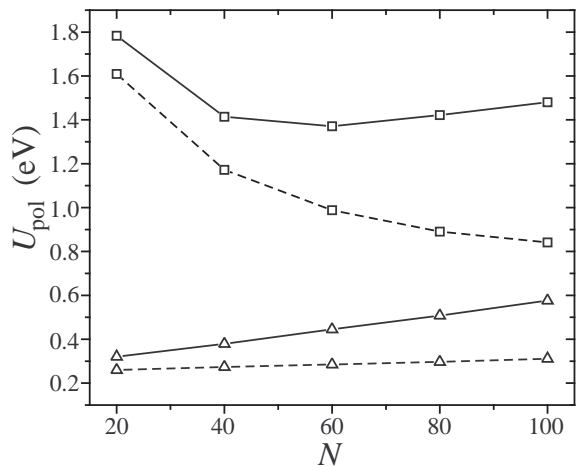}
\caption{The electrostatic energy stored in the solvent polarization as a function of the number $N$ of carbons in $\form$. Triangular data points are for neutral $\form^0$, squares for charged $\form^{-}$ chains. Dashed lines connect the data for RG1 chain geometry, solid lines for RG2.}
\label{fig:vover2}
\end{figure}

Bond-centric CDWs are known to form on neutral chains upon their dimerization with more electronic density corresponding to shorter ($\triple$) and less electronic density to longer ($\single$) bonds. Such CDWs have been widely discussed in the context of CPs as Peierls insulators \cite{YuLubook}. Being a non-vanishing distribution of the electric charge, a CDW is expected to polarize the solvent, which is indeed how we interpret the data shown in Fig.~\ref{fig:vover2}. The figure displays the electrostatic energy $\Upol$ stored in the solvent polarization as retrieved from outputs of Gaussian 03 computations. It is evident that this energy does not vanish for neutral chains and in fact grows linearly with the number of carbons $N$, as one would expect for a CDW. Since the amplitude of CDW is enhanced in neutral chains with RG2 geometry, they do exhibit a steeper growth with $N$.

Figure \ref{fig:vover2} also shows the evolution of $\Upol$  for negatively charged $\form^{-}$ chains. Especially illuminating here are the results for chains with RG2 geometry exhibiting a non-monotonic behavior of $\Upol$ as a function of $N$. $\Upol$ first decreases with $N$ at smaller $N$ reflecting the decrease of the electric field of the excess charge upon its spreading over longer oligomers. $\Upol$ then reaches its minimum in the region of $N$ corresponding to the formation of the self-consistent localized polaron pattern. With a further increase of $N$, $\Upol$ starts growing again with the same slope that is exhibited by neutral chains reflecting the interaction of the CDW with the solvent. The converged large-$N$ difference in $\Upol$ of charged and neutral chains with RG2 geometry is evidently quite substantial, close to 0.9 eV. While the minimum of $\Upol$ is not achieved for negative chains with RG1 geometry within the range of chain lengths studied, one could guess from the data of Fig.~\ref{fig:vover2} that the large-$N$ difference in $\Upol$ of charged and neutral chains in that case would probably be roughly half of the value for RG2 geometry. This ratio would look like more in line with charge localization patterns observed for RG1 and RG2. The very magnitude of the ``excess'' polarization energy just discussed may be indicative of the polaron binding energies appreciably larger than 0.1 eV. This speculative conclusion could be verified more directly if we were able to perform \textit{ab initio} computations for charged chains with the polarization state of the solvent corresponding to that for neutral chains.

\section{\label{BLA}Solvation and bond alternation patterns}

\begin{figure*}
\centering
\includegraphics[scale=0.7]{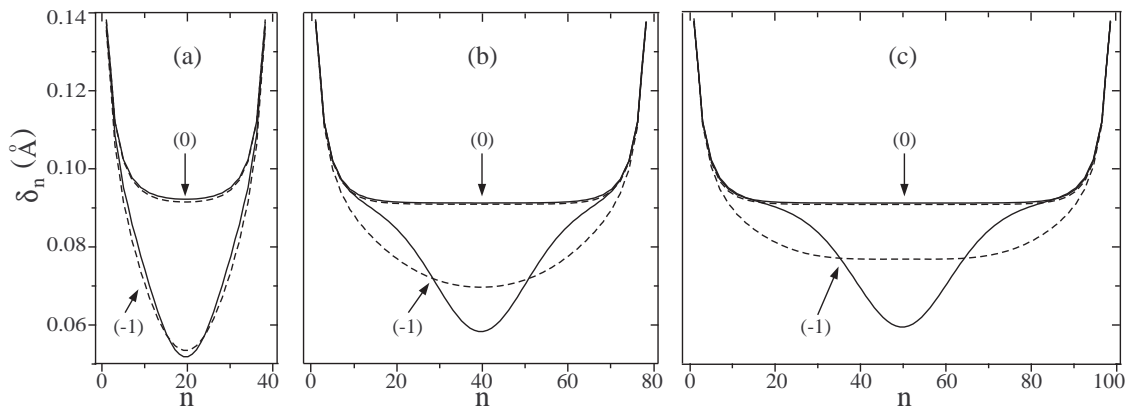}
\caption{Carbon-carbon bond-length alternation patterns for (a) $\cfourty$ (b) $\ceighty$ and (c) $\chundred$. Patterns for neutral chains are indicated with symbol (0), patterns for negatively charged chains with (-1). Dashed lines show results for chains in vacuum, solid lines for chains in the solvent environment.}
\label{BLA_combined}
\end{figure*}

We now turn our attention to the results obtained with full geometrical optimizations of underlying carbon lattices. A very convenient and well-known \cite{YuLubook,yang_kertesz_1,yang_kertesz_2} way to characterize dimerized polymeric structures is via bond-length alternation (BLA) patterns
$$
\dn = (-1)^n \, \left(l_n - l_{n+1} \right),
$$
where $l_n$ is the length of the $n$th carbon-carbon bond. In the infinite dimerized neutral structure, this pattern would be uniform, that is, independent of the spatial bond position $n$.

Figure \ref{BLA_combined} compares BLAs computed for neutral and negatively charged oligomers $\cfourty$, $\ceighty$ and $\chundred$. This comparison allows one to visually evaluate the magnitude of oligomer end effects. It is clear that a uniform, infinite-polymer, behavior is practically achieved over a long central-part stretch of the neutral $\chundred^{0}$. There is very little difference in BLAs of neutral chains in vacuum and in solvent; a small increase of $\dn$ in the solvent is consistent with our expectation for a CDW interacting with the environment.

Much more significant variations are observed for charged oligomers. As was discussed at length for electron-lattice polarons in CPs \cite{YuLubook,Bredas1985}, the magnitude of BLAs would exhibit a dip  in the spatial region containing an excess particle, where the lattice becomes ``less dimerized''. This expected decrease of $\dn$ is evident in all panels of Fig.~\ref{BLA_combined}. The character of its evolution with the oligomer length, however, is quite different for chains in vacuum and in solvent. The vacuum  data just shows a progressive reduction of the dip in $\dn$ for longer oligomers. In the absence of data for much longer chains, all we can conclude is that a very flat pattern of $\dn$ in the middle of $\chundred^{-}$ does not suggest yet an incipient formation of a self-localized pattern characteristic of polarons. This is reminiscent of the apparent failure of DFT computations to detect electron-lattice polarons in computationally tractable oligomers discussed in Sec.~\ref{Intro} -- with our results now extended to hybrid B3LYP computations. On the contrary, the data derived for chains in solvent quite clearly indicates a nearly self-consistent localized pattern of $\dn$ on $\chundred^{-}$, although, of course, longer oligomers would be needed to reveal a truly self-consistent polaron size.

\begin{figure}
\centering
\includegraphics[scale=0.6]{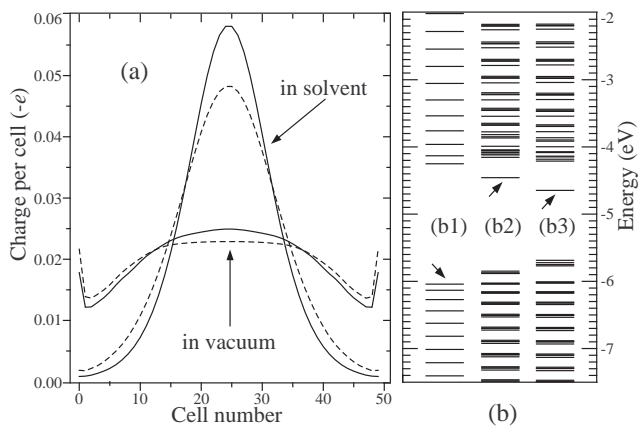}
\caption{Comparison of the solvation and BLA effects for $\chundred$. (a) Excess charge distributions in vacuum and in the solvent for the fully optimized lattice geometry (solid lines) and for the RG1 geometry (dashed lines). (b) The corresponding MO energy levels in the solvent for: (b1) Optimized neutral chain, (b2) Charged chain in RG1 geometry, (b3) Optimized charged chain. The arrows show HOMO levels.}
\label{fig:c100h2_normal_opt_dens_bands}
\end{figure}

These observations correlate well with the results obtained for excess charge densities on $\chundred^{-}$ that are displayed in panel (a) of Fig.~\ref{fig:c100h2_normal_opt_dens_bands}. The panel compares the excess density on chains with the uniform BLA of neutral chains (RG1 geometry) and on chains where the relaxation of the BLA due to the excess charge has taken place. One can see that while the relaxation of the BLAs does lead to spatial contractions of charge densities both in vacuum and in solvent, the delocalized (over the whole oligomer) nature of the excess charge remains intact in the vacuum case. Excess charge localization due to solvation, on the other hand, gets evidently enhanced by the localized lattice response seen in Fig.~\ref{BLA_combined}(c). One could probably say that solvation and lattice relaxation here act synergistically to affect excess charge self-trapping. With our B3LYP computations, the effect of the solvation on the excess charge distribution is evidently much more pronounced. A complementary information is provided by panel (b) of Fig.~\ref{fig:c100h2_normal_opt_dens_bands} showing the effects of solvation and BLA relaxation on the electronic structure of MO levels. Particularly relevant here are the intragap levels. Interestingly, for the energies of these levels effects due to each of the mechanisms appear similar in magnitude. This suggests that one has to be careful in evaluation of \textit{self-localization per se}  based just on the appearance of the MO levels in their overall structure. The spatial behavior of individual MOs as well as the total spatial distribution of the excess charge need to be properly analyzed.

\subsection{Hartree-Fock results}

As mentioned in the Introduction, in this paper we do not pursue an extensive comparison of results obtained with various \textit{ab initio} methods. To emphasize the level-of-theory dependence of at least the quantitative aspects, we, however, find it pertinent to provide a glimpse at the nature of results derived with Hartee-Fock (HF) computations. HF calculations for conjugated polymeric structures have been discussed \cite{yang_kertesz_1,yang_kertesz_2} to result both in overestimated energy gaps and magnitudes of BLAs. At the same time, unlike pure-DFT computations, they have also been shown to yield self-localized polaronic states for excess charge carriers \cite{bobbert,pitea_1,bredas_3,zuppi2003}. Our findings for polyynic chains agree well with these known trends.

\begin{figure}
\centering
\includegraphics[scale=0.48]{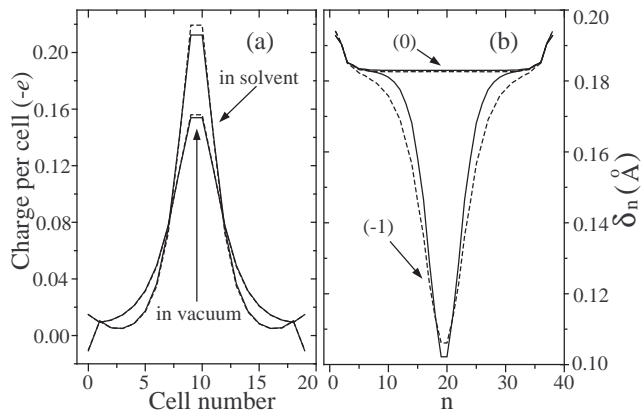}
\caption{Results of restricted-HF computations for $\cfourty$ oligomers. (a) Excess charge density distribution on $\cfourty^{-}$ in vacuum and in solvent. Dashed lines show the distribution on chains with a \textit{uniform} BLA pattern of the neutral system, solid lines are for chains with relaxed (optimal) geometries. (b) Optimal BLA patterns for neutral (0) and negatively charged (-1) oligomers. Dashed lines are for chains in vacuum, solid for chains in solvent.}
\label{fig:c40h2_HF}
\end{figure}

Figure \ref{fig:c40h2_HF} illustrates our restricted-HF results for neutral and negatively charged $\cfourty$ chains both in vacuum and in the solvent environment; they are consistent with a limited set of data we have also computed for longer chains. When compared to B3LYP-derived results discussed earlier in this paper and in Ref.~\cite{MGshort_prb}, one immediately notices an appreciably higher degree of excess charge density localization and larger magnitudes of BLAs obtained with the HF. Reassuringly, the qualitative effect of the solvent environment remains intact: it does enhance the excess charge localization (panel (a) of Fig.~\ref{fig:c40h2_HF}) and leads to a shorter and more pronounced associated dip in the optimized BLA pattern (panel (b)). Differently from the B3LYP-data, however, solvation here does not appear to act as a ``primary'' source.

In agreement with previous HF studies, we do find the excess charge self-localized already on chains in vacuum. One  might be tempted to say -- as expected for traditional electron-lattice polarons -- that the calculated  self-trapping in this case is \textit{caused} by the interaction with displacements of carbon atoms manifested in the relaxed BLA pattern in Fig.~\ref{fig:c40h2_HF}(b). The comparison made in Fig.~\ref{fig:c40h2_HF}(a), however, does not seem to support such a viewpoint. Specifically, we compare in that panel the excess charge densities obtained for fully relaxed geometries and for a rigid geometry featuring a \textit{uniform} BLA pattern. The latter corresponds to the BLA derived for neutral chains in their middle portion, Fig.~\ref{fig:c40h2_HF}(b). Despite the uniform lattice background, the excess charge localization evidently takes place, and with minimal variations from the fully geometrically relaxed chains. Moreover, the lattice relaxation, while indeed decreasing the total system energy, counter-intuitively results here in slightly smaller localization. (To compare, in B3LYP results geometrical relaxation would always lead to larger spatial charge localization.) Thus, it is rather the treatment of electron-electron interactions within the framework of HF computations, that appears to effect the excess charge localization observed in our HF data. In this paper, we forgo a discussion of the physical validity of these results that would require a more focused study.

\section{Discussion}

Self-localization of excess charge carriers (electrons or holes) into polaronic states on 1D semiconductor structures is an interesting effect that may be consequential for transport, optical, and electrochemical properties of these systems.  The basic nature of the self-trapping is easily revealed in a \textit{single-particle} picture within a standard 1D continuum adiabatic framework, where a localized carrier wave-function $\wfx$ is found as corresponding to the ground-state solution of the \textit{non-linear} Schr\"{o}dinger equation:
$$
-\frac{\hbar^2}{2m}\frac{\partial^2 \wfx}{\partial x^2} \,-\,
\int \! d\xp V(x-\xp)\, |\wfxp|^2 \,\wfx = E\wfx.
$$
Here $x$ is the coordinate along the structure aixs, $m$ the effective band  mass of the carrier, and $V(x)$ the effective self-interaction mediated by another subsystem. In the case of a short-range electron-phonon mediation, the self-interaction may be approximated as local: $V(x)=g\,\delta(x)$ leading then to the exact solution $\wfx \propto \mathrm{sech}(gmx/2\hbar^2)$ for a continuum 1D polaron. This is the result widely known after the
pioneering contributions of Rashba \cite{rashba_LP} and Holstein \cite{holstein_LP}. Another type of mediation arises due to the long-range polarization interaction of an excess charge carrier with a surrounding polar medium. In that situation, the effective $V(x)$ behaves as $1/|x|$ at large distances, while the short-range behavior depends on specific system details such as, e.g., the actual transverse charge density distribution and the geometry of the dielectric screening \cite{conwell_1,YNGpol,UG_optabs}. Numerical studies  of this case \cite{YNGpol,polcylinder} indicate that the binding energy of the resulting polarons could reach an appreciable fraction of the binding energy of the corresponding primary  optical excitations, Wannier-Mott excitons.

While illuminating the basic physics of the self-trapping, simplified single-particle descriptions have the drawback of not explicitly including valence-band electrons, whose interactions and reorganization may substantially affect the outcomes. Numerous \textit{ab initio} calculations have  been dedicated to effects of electron-electron interactions on electron-lattice polarons in CPs . In this paper we have extended our first model \textit{ab initio} study \cite{MGshort_prb} intended to address these effects in the formation of 1D polarons due to the interaction with a surrounding polar solvent. As our study is performed within the framework of the polarizable continuum model (PCM), it could also be thought of as a new avenue for applications of quantum-mechanical continuum solvation models \cite{leach_1,QMCSM}.

By studying and comparing various model systems based on long $\form$ oligomers with the polyynic structure, we have been able to successfully demonstrate the self-consistent formation of 1D large electron- and hole-polarons \textit{entirely} due to solvation. Reorganization of valence-band electrons and a more complex many-electron structure of the resulting polarons have been also explicitly illustrated. Full geometrical optimization of the underlying carbon lattice has shown that localized bond-length alternation patterns may act synergistically with the solvent reorganization to even further increase the degree of excess charge self-localization.

It should be reiterated that, while providing clear conceptual demonstrations and a seemingly appealing physical picture of solvation-induced self-localization, our current study is limited both in terms of its scope and direct quantitative applicability to real systems. Among other things, we have already discussed and illustrated a relatively strong dependence of the results on the level of \textit{ab initio} theory used, a common feature of such computations \cite{yang_kertesz_1,yang_kertesz_2}. DFT computations with hybrid exchange-correlation functionals may be ``appropriate'' engines for the problem at hand but a more comprehensive comparative study is required before conclusions can be drawn.  The PCM framework as used in this paper may overestimate the effects of solvation on self-localization as it does not take a proper account of the frequency dependence of the actual solvent dielectric functions \cite{YNGpol,polcylinder,polarons1,CTbook}. This issue could possibly be analyzed within newly developed time-dependent PCM schemes \cite{TimeDepPCM,SolvTDepend}. As the computational demand on calculations of 1D large polarons, especially in application to more complex semiconductor structures, may turn out to be excessive, one should also be looking for more efficient computational approaches. It would be interesting to see if the scaling description of solvent effects \cite{SolvScaling} may be applied to the problem of charge-carrier self-localization on 1D semiconductors.

\section{Acknowledgements}
We gratefully acknowledge financial support from the Collaborative U. T. Dallas -
SPRING Research and Nanotechnology Program.

\bibliography{abinitio_ref}

\end{document}